\documentclass[%
5preprint, 
 amsmath,amssymb,
 aps,
 prl,
]{revtex4-2}

\usepackage{graphicx}
\usepackage{dcolumn}
\usepackage{bm}
\usepackage{nicefrac}
\usepackage{array}

\usepackage{xcolor}

\begin{document}

\preprint{APS/123-QED}

\title{Limited coincidence between ultrahigh-field superconductivity and line of metamagnetic endpoints in UTe$_2$}

\date{\today}

\author{Peter Czajka$^{1,2}$}%
\thanks{These authors contributed equally: Peter Czajka and Sylvia K. Lewin}

\author{Sylvia K. Lewin$^{1,2}$}
\thanks{These authors contributed equally: Peter Czajka and Sylvia K. Lewin}

\author{Thomas Halloran$^{1,2}$}

\author{Corey E. Frank$^{1,2}$}

\author{Gicela Saucedo Salas$^2$}

\author{G. Timothy Noe, II $^3$}

\author{Sheng Ran$^4$}

\author{John Singleton$^3$}

\author{Nicholas P. Butch$^{1,2}$}
 
\affiliation{$^1$NIST Center for Neutron Research, National Institute of Standards and Technology, 
Gaithersburg, MD, USA}

\affiliation{$^2$Maryland Quantum Materials Center, Department of Physics, University of Maryland, College Park, MD, USA}

\affiliation{$^3$National High Magnetic Field Laboratory, Los Alamos National Laboratory, Los Alamos, NM, USA}

\affiliation{$^4$Department of Physics, Washington University in St. Louis, St. Louis, MO, USA}

\begin{abstract}
The field-dependent magnetization of UTe$_2$ was measured through the metamagnetic transition at a variety of field angles, tracking how the step in magnetization evolves with fields tilted away from the $b$ axis. For fields oriented within the $ab$ plane, jumps in both $M_a$ and $M_b$ vanish approximately 18° away from the $b$ axis. From contactless conductivity measurements, we find that the halo-like high-field superconducting region extends to the $ab$ plane, where it exists only within a very narrow ($<$1°) angular range near the termination of the metamagnetic phase boundary and extends beyond the highest measured field of 73~T. As the field orientation tilts towards the $c$ axis, the superconducting and metamagnetic phase boundaries no longer coincide and exhibit distinct trends.

\end{abstract}

\maketitle


\section{Introduction}

The heavy fermion superconductor UTe$_2$ exhibits likely spin-triplet pairing as well as a possible topologically nontrivial order parameter in its low-field superconducting phase. \cite{ran2019nearly,aoki2022unconventional,jiao2020chiral,hayes2021multicomponent} The material also has a complex phase diagram of magnetic field-induced states that have made the system a fruitful playground for high-magnetic-field science. \cite{ran2019extreme,miyake2019metamagnetic, lewin2023review} Of particular interest is UTe$_2$'s highest-field superconducting phase, which arises within a field-polarized (FP) magnetic state and which we thus denote SC$_{\rm{FP}}$; at ambient pressure, the lowest field at which this phase onsets is roughly 40~T. \cite{ran2019extreme,knebel2019field, knafo2021comparison} This is an enormous field scale relative to the phase's critical temperature of roughly 2~K. \cite{ran2019extreme, Helm2024compensation} In addition to this field scale, the SC$_{\rm{FP}}$ phase is remarkable for its field-angle dependence: it emerges with applied field \textbf{H} tilted off the crystallographic $b$ axis, creating a distinctive halo of high-field superconductivity in UTe$_2$ as a function of field angle. \cite{Lewin2024high}

As the FP state is effectively the parent state for the SC$_{\rm{FP}}$ phase, the nature of the FP state and its field-angle-dependence are of great interest. The FP state is separated from the low-field paramagnetic state by a metamagnetic transition whose associated transition field $H_{\rm{m}}$ is field-angle-dependent, generally increasing with field tilts away from $b$.\cite{miyake2019metamagnetic,knafo2019magnetic,knebel2019field,miyake2021enhancement,ran2019extreme,Lewin2024high} With $\mathbf{H} \parallel b$ and at low temperatures, the metamagnetic transition is first-order and occurs at $H_{\rm{m}} \approx$ 34~T; upon increasing temperature, the transition from the low-field paramagnetic state into the FP state changes from a first-order transition to a crossover at a critical endpoint (CEP) $T^* \approx$  7~K. \cite{miyake2019metamagnetic,miyake2021enhancement,knafo2019magnetic,Niu2020evidence,Thebault2022anisotropic}

The only angles at which the magnetization of UTe$_2$ across $H_m$ has been reported are $\mathbf{H} \parallel b$ and $\mathbf{H} \parallel [011]$; for both angles, the magnetization of UTe$_2$ jumps by $\approx$ 0.5 Bohr magnetons per formula unit at $H_{\rm{m}}$. \cite{miyake2019metamagnetic,miyake2021enhancement} Given the challenge of determining magnetic structure at fields 34~T and higher, magnetometry measurements of UTe$_2$ as a function of field angle are our most promising avenue to understanding the nature of the FP state.  In this work, we present angle-dependent magnetometry measurements that, when complemented by proximity detector oscillator (PDO) measurements, reveal new structure in UTe$_2$'s phase diagram that has substantial implications for the compound's exotic high-field superconductivity. 

\begin{figure*}
    \centering
    \includegraphics[width=\textwidth,trim={3cm 0.5cm 2cm 1.5cm},clip]{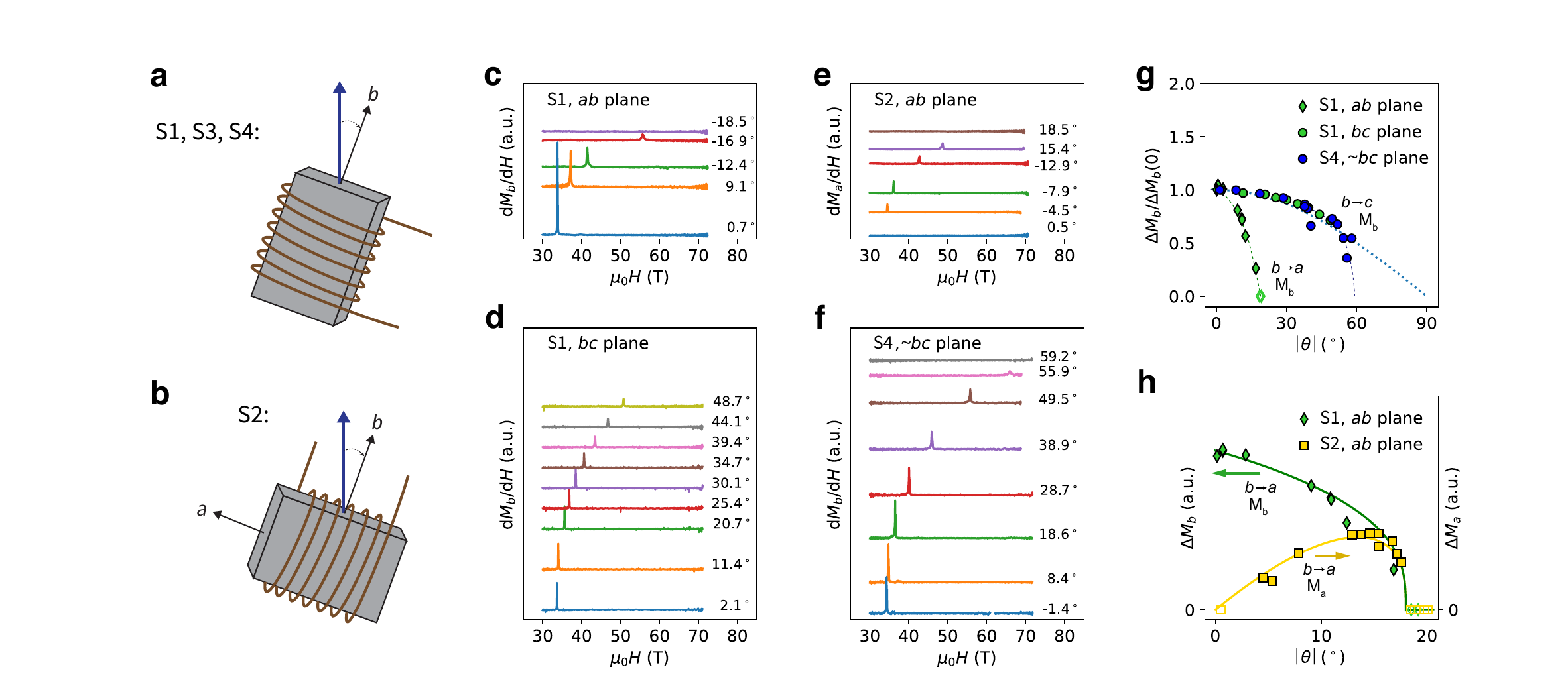}
    \caption{(a) Normalized magnetization jump along $b$ as a function of angle comparing field rotations in the $ab$ and $bc$ plane. Dashed lines are guides to the eye. Dotted line shows $\cos(\theta)$. (b) A comparison of magnetization jumps along $a$ and $b$ as a function of angle for rotations in the $ab$ plane. Dashed lines are guides to the eye. (c) $\mathrm{d}M_b/\mathrm{d}H$ versus field strength for sample S1 with field pulses performed at various angles within the $ab$ plane. (d) $\mathrm{d}M_b/\mathrm{d}H$ for sample S1 for fields within the $bc$ plane. (e) $\mathrm{d}M_a/\mathrm{d}H$ for sample S2 for fields within the $ab$ plane. (e) $\mathrm{d}M_b/\mathrm{d}H$ for sample S4 for fields within the $bc$ plane. For (c-f), all curve labels indicate the angle $\theta$ between the applied magnetic field and the $b$ axis, and all curves have been offset by an amount proportional to $\theta$ for visual clarity. (g) Illustration of the compensated coil  susceptometry measurement. In the case of samples S1, S3, and S4, the coil is oriented such that it measures the magnetization component along $b$ as the sample is rotated so that its $b$ axis is at an angle $\theta$ from the applied field. (h) In the case of sample S2, the coil is oriented such that it measures the magnetization component along $a$ as the sample is rotated so that its $b$ axis is at an angle $\theta$ from the applied field.}
    \label{fig:mag}
\end{figure*}

\section{Methods}

Single crystals of UTe$_2$ were grown using chemical vapor transport, with iodine as the transport agent. Table \ref{tab:samps} in Appendix A gives the masses of all samples used in this work as well as the conditions of their synthesis.

Measurements were performed using the 75~T duplex magnet system at the National High Magnetic Field Laboratory's Pulsed Field Facility at Los Alamos National Laboratory. We use the same procedure as Ref. \cite{Lewin2024high} for achieving effective dual axis rotation. The crystal is glued down on a single axis rotator platform such that \textbf{H} will lie in the $bc$ plane at some fixed angle $\theta_{bc}$ from $b$ to $c$. Rotation then permits a variable field component $\theta_a$ outside the $bc$ plane. For rotations within the $bc$ plane itself we instead simply mount the sample to rotate around the $a$ axis.

We use a compensated coil technique for magnetometry, as described in Ref. \cite{Goddard2008Coils}. The voltage induced in a coil around the sample is directly proportional to the time derivative of the sample's magnetization along the coil: $V = \alpha \times \mathrm{d}M_i /\mathrm{d}t$, where $\alpha$ is a constant depending on both the sample mass and the coil characteristics. Here $i$ represents the coil axis, which is aligned with the UTe$_2$ $b$ axis for the samples we call S1, S3, and S4 and aligned with the $a$ axis for the sample we call S2 (see Fig. \ref{fig:mag}a-b). The direction of the applied field is fixed; to measure at different field angles, the coil and sample are rotated together in situ between field pulses. Thus, we only measure the magnetization component along $i$. \textcolor{black}{During each pulse, the time derivative of the applied magnetic field $\mathrm{d}H/\mathrm{d}t$ is measured by a separate coil.} Dividing the induced voltage in the coil by $\mathrm{d}H/\mathrm{d}t$ yields the differential susceptibility, $\mathrm{d}M_i/\mathrm{d}H$, up to a constant of proportionality.

The differential nature of this method provides strong sensitivity to the metamagnetic transition, whose associated jump in magnetization appears as a sharp peak in $\mathrm{d}M_i/\mathrm{d}H$ at $H_m$. The size of the corresponding jump $\Delta M_i$ can be obtained by integrating the differential signal: $\Delta M_i = \int_{H_-}^{H_+} (\mathrm{d}M_i/\mathrm{d}H) \mathrm{d}H$ where $H_{-}$ and $H_{+}$ are fields on either side of the observed peak (details in Appendix E). 

We also performed PDO measurements, a contactless technique in which the sample is coupled inductively to an LC circuit. The resonant frequency of the circuit is then highly sensitive to changes in sample resistance and magnetization. \cite{altarawneh2009proximity} Traditional transport measurements of UTe$_2$ are difficult at cryogenic temperatures in pulsed fields \textcolor{black}{due to low resistivity, which leads to a low signal-to-noise ratio}. PDO measurements are a highly effective tool for studying phase diagrams in pulsed field environments \cite{Ghannadzadeh2011PDO}. Furthermore, the PDO response of UTe$_2$ \textcolor{black}{at field-induced phase transitions} is strong and well-understood. \cite{ran2019extreme,Lewin2024high,wu2023enhanced,weinberger2024quantum,Wu2025quantum} In previous PDO measurements, we called the measured samples P1 and P2. \cite{Lewin2024high} Here, we report additional measurements on P2 as well as measurements on a new sample, P3. The resonance frequency of our PDO circuit is typically in the range of 28 to 30~MHz, \textcolor{black}{above the sampling rate of our digitizer}. We mix the signal down to approximately 2~MHz using a double-heterodyne technique.

\section{Magnetometry}

Fig. \ref{fig:mag}c shows the magnetometry of sample S1 for a set of rotation angles $\theta_a$ within the $ab$ plane. The peak in $\mathrm{d}M_{b}/\mathrm{d}H$ shifts to higher $H$ as $\theta$ increases, consistent with the previously reported angle-dependence of $H_{\rm{m}}$. \cite{ran2019extreme,Schonemann2023MCE,Lewin2024high} Fig. \ref{fig:mag}g shows the extracted jump in $b$-axis magnetization, $\Delta M_b$, as a function of angle for rotations in both the $ab$ and $bc$ planes. The data are rescaled as $\Delta M_b/\Delta M_b^0$, where $\Delta M_b^0$ is $\Delta M_b (\theta = 0)$. This is done to normalize the data sets obtained for S1 in the two rotational planes (\textcolor{black}{which differ in constant of proportionality due to sample position within the coil between the two measurements}) and for sample S4.

It is clear that $\Delta M_b$ decreases with increasing field angle away from $b$, both in the $ab$ and $bc$ planes, but that rotations toward $a$ suppress the metamagnetic transition more dramatically.

In the $ab$ plane, $\Delta M_b$ decreases rapidly with increasing $\theta_a$ and goes to zero at some critical angle $\theta_a^{crit} \approx$~18°. Correspondingly, no peak is visible in $\mathrm{d}M_b/\mathrm{d}H$ for $\theta_a \geq$~18.5° (top curve in Fig. \ref{fig:mag}c). \textcolor{black}{Note that the metamagnetic transition occurs at $\approx$~55~T at $\theta_a \approx$~17°. Based on the evolution of $H_m$ with $\theta_a$, $H_m$ should still be well below the maximum measurement field of 73~T at $\theta_a \approx$~18°.} This suggests that there is a critical angle at which the metamagnetic transition disappears or ceases to be discontinuous.

In the case of field rotations in the $bc$ plane, $\Delta M_b$ decreases with $\theta_{bc}$, but at a slower rate, consistent with the \textbf{H} $\parallel$ [011] measurement of Ref. \cite{miyake2021enhancement}. For sample S4, shown in Fig. \ref{fig:mag}f, features in $\mathrm{d}M_b/\mathrm{d}H$ were observed up to 73~T, the highest field at which measurements were performed (see Fig. \ref{fig:appInt} in Appendix E). As detailed in Appendix F, analysis of the rate at which $H_m$ increases with angle indicates that S4 was actually rotated by approximately 10° from the $bc$ plane when measured. Based on the data in/near the $bc$ plane, it is plausible that $\Delta M_b$ is proportional to $\cos(\theta_{bc})$ in the $bc$ plane; it is also plausible that it drops off similarly to the $ab$ plane but with $\theta_{bc}^{crit} \approx$~60 to 65°, as suggested by the linear extrapolation in Ref. \cite{Wu2025quantum}. 

The susceptibility tensor of UTe$_2$ must be diagonal based on its orthorhombic crystal symmetry \cite{Nye1985}. As expected, there is no feature in $\mathrm{d}M_a/\mathrm{d}H$ with field along $b$, as seen in the lowest curve of Fig. \ref{fig:mag}e. With increasing tilts from $b$ towards $a$ there is first a rise and then a decrease in $\Delta M_a$. Just as with the $b$-axis magnetization, there is no feature in the measured $a$-axis magnetization for $\theta_a \geq$~18.5 in the $ab$ plane. Therefore, the discontinuity in total magnetization at $H_m$ has disappeared by $\theta_a^{crit} \approx$~18°, not merely the $b$-axis component.

We also measured the $c$-axis magnetization of a separate sample in the $bc$ plane and did not observe a jump in $M_c$ at $H_m$ (see Fig. \ref{fig:appM6} in Appendix E). This is consistent with torque magnetometry measurements that showed a non-zero torque in the FP phase for fields in the $bc$ plane, showing that the magnetization and applied field are not collinear. \cite{Helm2024compensation} It appears that for fields in the $bc$ plane, the jump in magnetization at $H_m$ is purely along the $b$ axis. From the $1/\cos(\theta_{bc})$ dependence of $H_m$, it can be concluded that only the $b$-axis component of field causes the metamagnetic transition for fields in the $bc$ plane. \cite{Lewin2024high} It is consistent that for fields in this plane, only the $b$-axis component of magnetization is affected at $H_m$.

On the other hand, in the $ab$ plane the $a$-axis magnetization is also affected at the metamagnetic transition. If the jump in magnetization is collinear with field direction in the $ab$ plane, then it follows that
\begin{equation}\label{eq:Deltas}
\begin{split}
\Delta M_a (\theta) &= \Delta M (\theta) \sin(\theta),\\
\Delta M_b (\theta) &= \Delta M (\theta) \cos(\theta),
\end{split}
\end{equation}
 where $\Delta M (\theta)$ indicates the overall magnitude of the increase in magnetization at $H_m$. We only have measurements of both $\Delta M_a$ and $\Delta M_b$ up to constants of proportionality. So we use a modified version of Eq. \ref{eq:Deltas}:
 \begin{equation}\label{eq:Deltas2}
\begin{split}
\Delta M_a (\theta) &= c_a\Delta M (\theta) \sin(\theta),\\
\Delta M_b (\theta) &= c_b\Delta M (\theta) \cos(\theta),
\end{split}
\end{equation}
where $c_a$ and $c_b$ are constants.
The $ab$ plane data for angles below $\theta_a^{crit}$ can be well-described by Eq. \ref{eq:Deltas2}, as shown in Fig. \ref{fig:mag}h. For these fits, we used $\Delta M (\theta) = (\frac{\theta_a^{crit} - \theta}{\theta_a^{crit}})^{1/3}$, \textcolor{black}{a relation that resembles an order parameter, with the exponent chosen as a simple fraction that fits the data well}. The applicability of Eq. \ref{eq:Deltas2} indicates that in the $ab$ plane, the direction of the jump in magnetization at $H_m$ is collinear with the direction of the applied magnetic field. Thus, we conclude that the magnetic moment in the FP phase is constrained to the $ab$ plane, within which it can rotate to align with the applied field.

\begin{figure}[h!]
    \centering
    \includegraphics[width=0.5\textwidth,trim={1.5cm 0cm -1cm 1cm},clip]{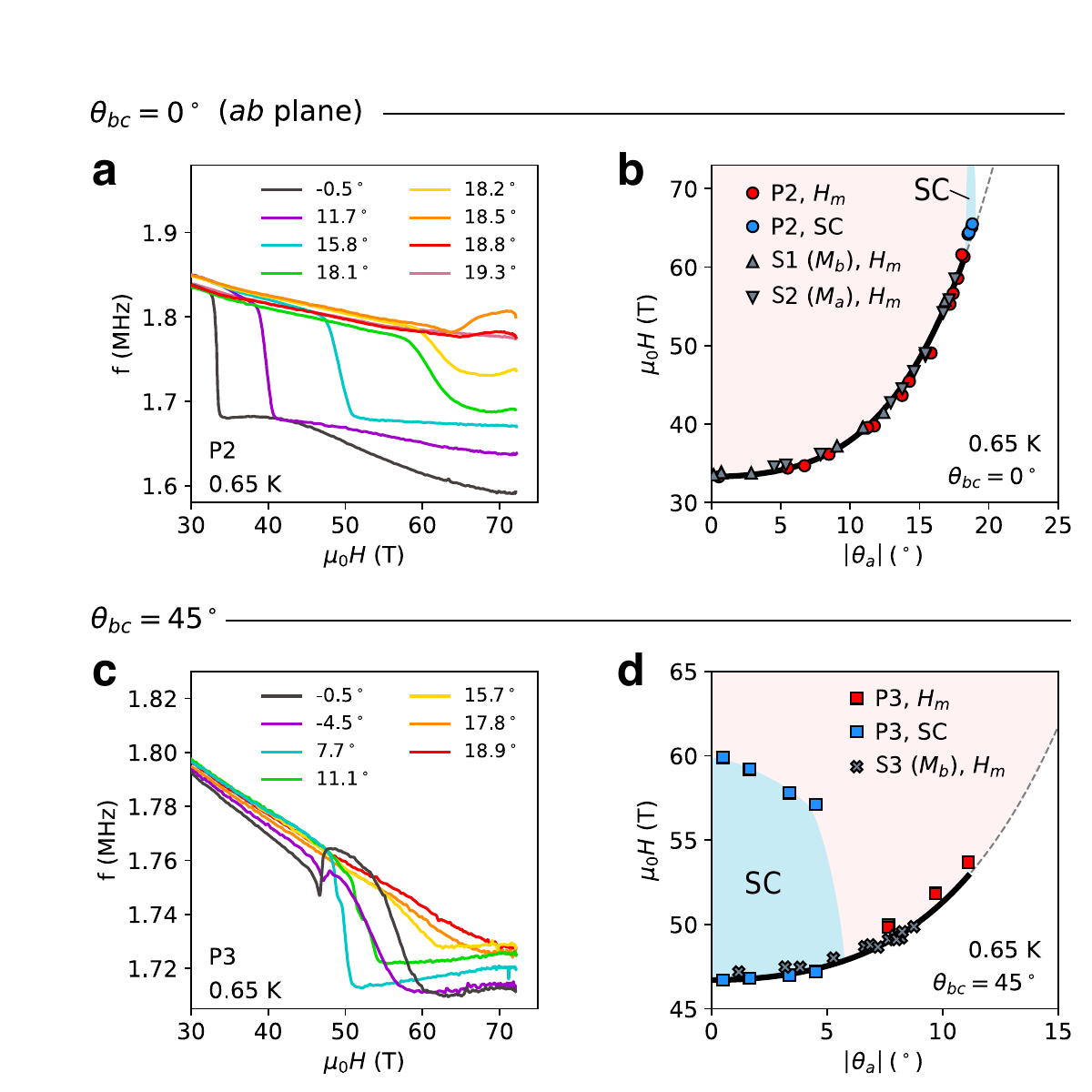}
    \caption{(a) PDO frequency as a function of field strength for pulses performed at various angles $\theta_a$ within the $ab$ plane. (b) Field angle-dependent phase diagram for fields applied within the $ab$ plane. (c) PDO frequency as a function of field strength for pulses performed at various angles $\theta_a$ away from the $bc$ plane, at fixed $\theta_{bc} =$~45°. (d) Field angle-dependent phase diagram for fields applied at various angles $\theta_a$ away from the $bc$ plane, at fixed $\theta_{bc} =$~45°. In (b) and (d), gray points indicate angles at which a metamagnetic transition is observed in magnetometry; blue (red) points from PDO samples indicate angles at which a  superconducting transition (step-like metamagnetic transition) is observed. }
    \label{fig:slices}
\end{figure}

\section{Proximity detector oscillator measurements}

The change in PDO frequency is opposite in sign to a change in resistivity of the sample. Thus, in UTe$_2$ the PDO frequency $f$ abruptly decreases when the sample enters the FP state or increases when the sample enters the SC$_{\rm{FP}}$ state. \cite{ran2019extreme,Lewin2024high,wu2023enhanced,weinberger2024quantum,Wu2025quantum} (Note that the results in Ref. \cite{Wu2025quantum} appear to differ from this by a sign change, due to their definition of $\Delta f$.)

Consistent with magnetometry, PDO measurements indicate a suppression of the metamagnetic transition with rotations of \textbf{H} from $b$ towards $a$ (see Fig. \ref{fig:slices}a), as previously reported in Ref. \cite{Wu2025quantum}. As shown for sample P2 in Fig. \ref{fig:slices}a, a jump downward in $f$ is observed for \textbf{H} directed close to $b$. As the field is tilted from $b$ towards $a$, the transition softens considerably.

At $\theta_a =$~18.5°, we observe an increase in $f$ at $\approx$~64~T. The identification of this feature with superconductivity is supported by its disappearance at higher temperature, as shown in Fig. \ref{fig:temps}a. Similar behavior is found at 18.6° (not shown) and 18.8°. For 19.3°, $f$ vs. $H$ is featureless, exhibiting neither a metamagnetic nor a superconducting transition.

These measurements demonstrate that the halo-shaped SC$_{\rm{FP}}$ region identified by Lewin \textit{et al.} \cite{Lewin2024high} extends all the way to the $ab$ plane, forming a continuous ring around the $b$ axis. The narrow angular extent ($<$1°) and the high onset field ($\approx$~65~T) of SC$_{\rm{FP}}$ in the $ab$ plane are remarkable. The form of the $f$ vs. $H$ curves also suggests that the upper critical field in the $ab$ plane well exceeds 73~T. \textcolor{black}{In layered organic superconductors, field-induced superconductivity has been observed that only exists for fields directed parallel or near-parallel to the conducting layers \cite{UjiOrganic2001,KonoikeOrganic2004,balicas2001superconductivity}. This field-angle limitation is due to the highly anisotropic orbital effect in these quasi-2D materials. We are unaware of any superconductor besides UTe$_2$ where such extreme angle sensitivity is observed but the field angle that induces superconductivity is not related to a high symmetry plane.} Of note, the narrow window of superconductivity in the $ab$ plane corresponds almost exactly to $\theta_a^{crit}$ at which the step in magnetization at $H_m$ disappears, as shown in the phase diagram in Fig. \ref{fig:slices}b.

In contrast, consider a different part of UTe$_2$'s field-angle diagram: with field tilted 45° from the $b$ axis towards $c$, i.e. $\theta_{bc} =$~45°, UTe$_2$ enters the SC$_{\rm{FP}}$ phase at $H_m$. This can be seen by the sharp jump in $f$ of sample P3 in the black curve in Fig. \ref{fig:slices}c. When the sample is rotated so that the field is not purely in the $bc$ plane, the superconducting field region first narrows and then disappears at $\theta_a \approx$~5°. However, for higher $\theta_a$ there is a sharp jump downwards in the PDO frequency until at least $\theta_a \approx$~11°. Similarly, features in the magnetization of sample S3 can be observed up to $\theta_a \approx$~11° (see Fig. \ref{fig:appS3} in Appendix E). The phase diagram for $\theta_{bc} =$~45° is shown in Fig. \ref{fig:slices}d. At this value of $\theta_{bc}$, superconductivity appears in the field-angle region of a strong first-order metamagnetic transition.

\begin{figure}[]
    \centering
    \includegraphics[width=0.5\textwidth,trim={0.5cm 7cm 0.5cm 5.5cm},clip]{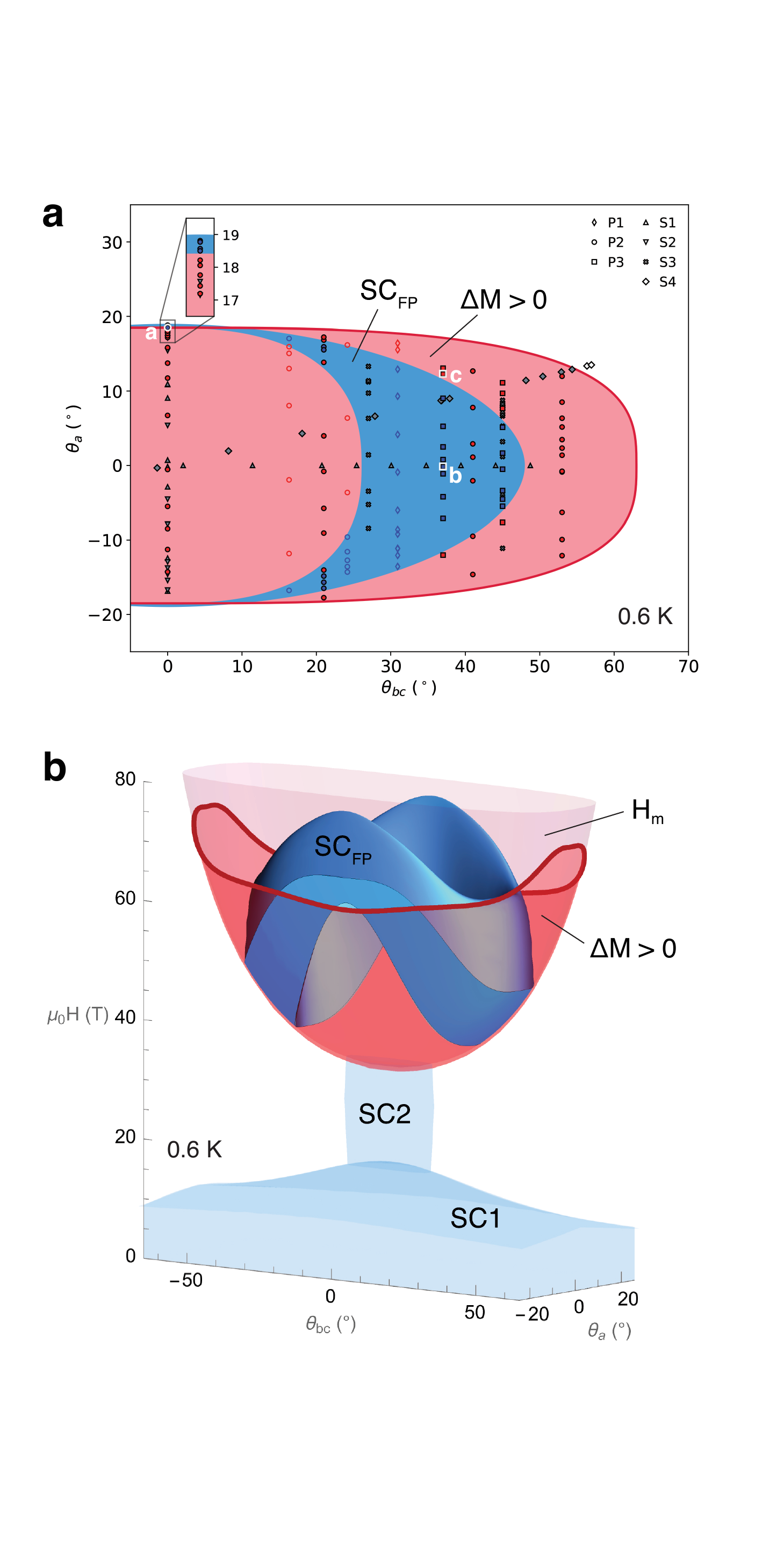}
    \caption{(a) High-field phase diagram for UTe$_2$ at 0.6 K as a function of $\theta_a$ and $\theta_{bc}$ based on magnetization and PDO measurements. PDO data shown are a compilation of PDO measurements from this report (samples P2 and P3; filled symbols) and from Ref. \cite{Lewin2024high} (samples P1 and P2; empty symbols). Blue (red) points indicate angles where a superconducting transition (step-like metamagnetic transition) occurs. Gray points indicate angles at which a peak in $\mathrm{d}M_i/\mathrm{d}H$ was observed in magnetometry. Open black symbols indicate angles at which a partial peak was observed but was cut off at our maximum measurement field of 73~T. Blue and red shaded regions are guides to the eye for the SC$_{\rm{FP}}$ region and the $\Delta M > 0$ region described in the text. Inset is a zoom-in on select points in the $ab$ plane. Data points highlighted in white indicate angles at which temperature-dependent PDO measurements were performed; the labels a, b, and c correspond to the subfigures of Fig. \ref{fig:temps}. (b) Illustration of the 0.6 K phase diagram of UTe$_2$ as a function of $\theta_a$, $\theta_{bc}$, and magnetic field strength, based on the measurements shown in (a) as well as Ref. \cite{LewinFieldangle2024, Lewin2024high}. Away from the $ab$ plane, the boundary of the $\Delta M > 0$ region does not coincide with the field angles at which the SC$_{\rm{FP}}$ phase exists.}
    \label{fig:phasediag}
\end{figure}

The overall field-angle phase diagram for UTe$_2$ at high fields is illustrated in Fig. \ref{fig:phasediag}(a), which presents the high-field behavior as a function of $\theta_{bc}$ and $\theta_a$, disregarding field strength. Blue (red) points are used for angles where a superconducting (metamagnetic) transition occurs. Included are only metamagnetic transitions that show a clear step in PDO frequency; as an example, in Fig. \ref{fig:slices}c there is a clear step in $f$ for $\theta_a =$~11.1° but only a kink for $\theta_a =$~15.7°. Magnetometry results are also shown: gray points indicate angles at which a peak in $\mathrm{d}M/\mathrm{d}H$ was observed. Blue and red shaded regions are intended as guides to the eye, showing the SC$_{\rm{FP}}$ region and the region in which there is a step in PDO frequency at $H_m$, which we label ``$\Delta M > 0$."

The general picture that emerges is of a superconducting halo around the $b$ axis that becomes thin in angular extent as \textbf{H} approaches the $ab$ plane. The field angle at which superconductivity emerges in the $ab$ plane is near the edge of the $\Delta M > 0$ region.  In contrast, our study and Ref. \cite{Wu2025quantum} show that the $\Delta M > 0$ region extends to at least $\theta_{bc}$ = 60° in the $bc$ plane, while the upper boundary of the SC$_{\rm{FP}}$ halo is around $\theta_{bc} = $~45° at 0.6~K \cite{ran2019extreme,Wu2025quantum}. It is observed that for each PDO sample, there are values of $\theta_{bc}$ for which the $\Delta M > 0$ region extends beyond SC$_{\rm{FP}}$ as a function of $\theta_a$.

Note that there is some variability associated with individual sample characteristics. For example, sample P2 has a mass of 1.4~mg, over three times larger than P3's mass of 0.4~mg. Therefore, P2 may experience more eddy current heating during each magnetic field pulse than P3. This can bring the sample above the critical temperature of the SC$_{\rm{FP}}$ phase. In Fig. \ref{fig:phasediag}a it is apparent that sample P2 is not superconducting at $\theta_{bc} \approx$~40°, in a region of field angles where P3 is superconducting.

The field-angular boundary of the SC$_{\rm{FP}}$ phase is notable, considering the magnetic and electronic anisotropy in UTe$_2$. For fields below $H_m$ and at low temperatures, the $a$ axis is the magnetic easy axis of UTe$_2$, with magnetic susceptibility several times higher than those of the $b$ or $c$ axes. \cite{ran2019nearly} From the magnetometry results described above, it is clear that there is also magnetic anisotropy in the FP phase, although its character is different. Magnetization along $a$ is affected at the metamagnetic transition while magnetization along $c$ is not, and $\Delta M_b$ evolves very differently with tilts from $b$ towards $a$ versus towards $c$. Despite this magnetic anisotropy, the SC$_{\rm{FP}}$ halo has a nearly circular inner diameter of approximately 20° tilt away from the $b$ axis. This near-symmetry about the $b$ axis differs dramatically from the $c$-axis alignment of UTe$_2$'s cylindrical Fermi surfaces at low field.\cite{Aoki2022first,Eaton2024quasi,broyles2023revealing}

\section{Relationship between metamagnetic discontinuity and superconductivity}

It has been proposed that the 7~K metamagnetic CEP at $\mathbf{H} \parallel b$ is suppressed to lower temperatures as field is tilted away from $b$, eventually reaching 0~K and thus becoming a quantum critical endpoint (QCEP). \cite{Wu2025quantum} At present, there is only one system (Sr$_3$Ru$_2$O$_7$) for which a field angle-tunable metamagnetic QCEP has been reported.\cite{grigera2001magnetic,grigera2003angular} Notably, the proliferation of quantum fluctuations at a quantum critical point can stabilize superconductivity. \cite{gegenwart2008quantum,abrahams2011quantum,park2006hidden,coleman2005quantum} In that case, superconductivity should be strongest at the QCEP. From our measurements, we can assess whether QCEPs play a role in the SC$_{\rm{FP}}$ phase of UTe$_2$.

In an ideal system, as a first-order transition is suppressed, the jump in order parameter should remain infinitesimally sharp and merely decrease in magnitude. However, in our measurements and those reported in Ref. \cite{Wu2025quantum}, the jump in PDO at $H_m$ broadens fairly continuously with increasing field tilt from $b$; a similar trend appears in our magnetization data (see Fig. \ref{fig:appIntShift} in Appendix E). Due to this discrepancy between prediction and observation, it can be difficult to pinpoint the exact angle at which a first-order transition becomes a smooth crossover. Previously, the temperature of the CEP for $\mathbf{H} \parallel b$ in UTe$_2$ was mainly determined by the disappearance of magnetic hysteresis in measurements of magnetization, magnetoresistance, and thermoelectric power.  \cite{miyake2019metamagnetic,knafo2019magnetic,Thebault2022anisotropic,Niu2020evidence} Previous resistance measurements in DC fields show that the size of the hysteresis loop at $H_m$ is roughly unchanged up to $\theta_a =$~8° in the $ab$ plane and up to $\theta_{bc} =$~24° in the $bc$ plane. \cite{LewinFieldangle2024} As discussed in Appendix E, analyzing hysteresis was inconclusive for these measurements. Therefore, we differentiate between PDO and magnetization signals that show a clear step as a function of field vs. those that do not, at a measurement temperature of $\approx$~0.6~K (see dark red line in Fig. \ref{fig:phasediag}b).

As shown in Fig. \ref{fig:slices}, there are opposite trends for $\theta_{bc} =$~0° vs. $\theta_{bc} =$~45° in terms of the relation between SC$_{\rm{FP}}$ and the jump in magnetization at $H_m$. In both cases, the downward jump in $f$ at $H_m$ becomes softer and broader as $\theta_a$ increases, as shown in Fig. \ref{fig:slices}a and Fig. \ref{fig:slices}c. PDO measurements performed at $\theta_{bc}$ = 23°, 41°, 52° for P2 and at 35° for P3 reveal a similar angle dependence: the negative step in the PDO signal associated with metamagnetism shows substantial softening with increasing $\theta_a$. However, the field-angle-dependence of the SC$_{\rm{FP}}$ phase is very different to that of the metamagnetic transition. For $\theta_{bc} =$~0°, i.e., the $ab$ plane, superconductivity arises  approximately at $\theta_a^{crit}$, the angle where the feature in magnetization disappears altogether. Yet at $\theta_{bc} =$~45°, superconductivity exists at $\theta_a =$~0° and disappears at higher $\theta_a$. In other words, for higher $\theta_{bc}$, superconductivity exists where the jump in magnetization is largest as a function of $\theta_a$.

\begin{figure}[h!]
    \centering
    \includegraphics[width=0.5\textwidth,trim={0.5cm 0cm -1cm 0cm},clip]{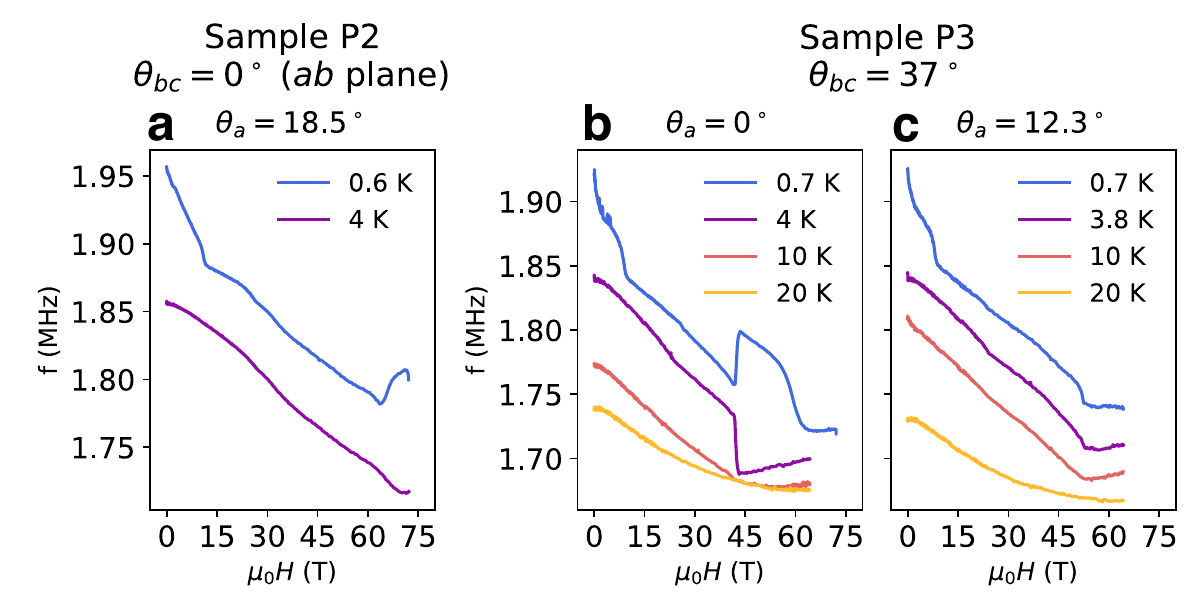}
    \caption{Temperature-dependent PDO measurements of (a) sample P2 at $\theta_{bc}$~=~0° and $\theta_a$~=~18.5°; (b) sample P3 at $\theta_{bc}$~=~35° and $\theta_a$~=~0$\pm$0.2°; and (c)  sample P3 at $\theta_{bc}$~=~35° and $\theta_a$~=~12.3$\pm$0.1°. The data points highlighted in white in Fig. \ref{fig:phasediag}a indicate the field angles at which these measurements were performed, in context of the overall phase diagram.}
    \label{fig:temps}
\end{figure}

The coexistence of superconductivity with strong metamagnetic transitions can be seen clearly by looking at the temperature dependence of the PDO data at fixed field angle. Fig. \ref{fig:temps}b shows the PDO signal of sample P3 with field in the $bc$ plane (i.e., $\theta_a$~=~0°) at $\theta_{bc}$~=~35°. At 0.7~K, there is a sharp jump in the PDO frequency at roughly 42~T, indicating the onset of the SC$_{\rm{FP}}$ phase. In the normal state at 4~K, there is a similarly sharp drop in PDO frequency, showing a clear first-order metamagnetic transition. Compare this condition to an orientation approximately 12° outside the $bc$ plane, as shown in Fig. \ref{fig:temps}c, where there is no superconductivity but only a weak metamagnetic transition. It is clear that even at the lowest measured temperature of 0.7~K, the jump in magnetization at this field angle is smaller than the jump at 4~K for field at $\theta_a$~=~0°, i.e., in the $bc$ plane. This suggests that the CEP is at a higher temperature for $\theta_a$~=~0° than for $\theta_a$~=~12°. However, the SC$_{\rm{FP}}$ phase exists at $\theta_a$~=~0°, where the CEP is at higher temperature. This correlation is opposite what would be expected if the SC$_{\rm{FP}}$ phase were due to quantum fluctuations at field angles where the metamagnetic critical endpoint is suppressed to 0~K.

\section{Discussion}
 We find that the increase in magnetization at the metamagnetic transition, $\Delta M$, is largest for fields along $b$ and decreases as field is tilted away from $b$. Field tilts in the $ab$ plane suppress $\Delta M$ the most efficiently, with no measurable jump in magnetization observed beyond a critical angle of $\theta_a^{crit} \approx$~18°. The jump in magnetization is confined to the $b$ axis for fields in the $bc$ plane. However, we find evidence that $\Delta M$ is collinear with the direction of the applied field for fields in the $ab$ plane. Future torque magnetometry measurements in the $ab$ plane of UTe$_2$ could be used to independently verify this conclusion.

Through PDO measurements, we have found that the halo-shaped SC$_{\rm{FP}}$ phase does in fact extend all the way to the $ab$ plane, but only within a very narrow angular window. This window notably occurs roughly at $\theta_a^{crit}$, indicating that there may be an interplay between the suppression of metamagnetism and high-field superconductivity in the $ab$ plane. 

One possible explanation is that the boundary of the FP phase itself is at $\theta_a \approx$~20° in the $ab$ plane. Since the SC$_{\rm{FP}}$ phase can only exist in the FP phase, $H_m$ serves as a lower field bound to the SC$_{\rm{FP}}$ phase \cite{ran2021expansion}. Similarly, the boundary of the FP phase in terms of field angle would be expected to limit the extent of the SC$_{\rm{FP}}$ phase. Previous measurements have found a non-monotonic temperature dependence of $H_m$ at $\theta_a =$~18° in the $ab$ plane, suggesting a possible competition between nearly degenerate magnetic ground states.\cite{Wu2025ripples} Perhaps the FP phase that can host superconductivity disappears at $\theta_a^{crit}$ and is replaced by a different electronic state.

In Ref. \cite{Wu2025quantum}, the extension of SC$_{\rm{FP}}$ to magnetic fields below $H_m$ in resistance measurements was cited as evidence of quantum-criticality-induced superconductivity, as it occurred for field angles near the putative QCEPs. However, we note that a similar appearance of the SC$_{\rm{FP}}$ phase below $H_m$ was observed in resistance measurements under pressure; in these measurements the metamagnetic transition was clearly first-order, as demonstrated by the presence of hysteresis. \cite{ran2021expansion} Rather than arising from proximity to a QCEP, we posit that such ``spillover" may be due to FP domain formation below $H_m$, which could result in filamentary superconductivity.

For field orientations outside the $ab$ plane, we find no correspondence between the existence of the SC$_{\rm{FP}}$ phase and the end of the first-order metamagnetic phase transition. This indicates that quantum critical fluctuations originating from the metamagnetic phase boundary are not the source of superconducting pairing for the SC$_{\rm{FP}}$ phase.

Several hypotheses remain for the unusual field-angle-dependence of UTe$_2$'s SC$_{\rm{FP}}$ phase. While our results do not support the connection between QCEPs and the SC$_{\rm{FP}}$ phase, they do not preclude the general possibility that the superconducting pairing strength is field-angle-dependent. In this case, the superconducting transition temperature $T_c$ should also vary with field angle. Another possibility is that the Cooper pairs carry a finite angular momentum, requiring a non-unitary spin-triplet order parameter. \cite{Lewin2024high} Then the upper critical field depends on field-angle due to a coupling between field and the superconducting order parameter. A separate theory that has been explored is that the SC$_{\rm{FP}}$ phase involves interband pairing, made possible by Zeeman splitting. \cite{Yu2025Pauli} Within this framework, the field-angle dependence of the superconductivity arises from spin-orbit coupling.

Future experiments will be needed to map the  upper critical field of the SC$_{\rm{FP}}$ phase as a function of field angle and temperature. The highest upper critical field of the SC$_{\rm{FP}}$ phase is currently unknown; for many field angles, it exceeds 73~T, our maximum applied field in these measurements. Such a study may help to distinguish between various theories of the SC$_{\rm{FP}}$ phase. It will also be valuable to further study the FP phase itself, building on our extant magnetometry measurements to understand the magnetization vector of UTe$_2$ across the entire FP phase.

\section{Conclusion}

The highest-field superconducting phase of UTe$_2$, the SC$_{\rm{FP}}$ phase, has a unique field-angle dependence among all known field-induced superconductors. We find that the SC$_{\rm{FP}}$ phase extends all the way to the $ab$ plane of UTe$_2$, forming a complete halo around the $b$ axis in terms of field angle. We also find that this angular dependence does not appear connected to the endpoints of the first-order metamagnetic transition into the FP phase. Regarding the FP phase that hosts high-field superconductivity, we find evidence that the jump in magnetization at the metamagnetic transition of UTe$_2$ is collinear with field direction for fields in the $ab$ plane. This contrasts with the purely $b$-axis jump in magnetization we find for fields in the $bc$ plane. All of these observations provide much desired footing for microscopic theories and firm constraints for testing future models.

\section{Acknowledgments}
We thank Andriy Nevidomskyy for helpful discussions. This work was supported in part by the National Science Foundation under the Division of Materials Research Grant NSF-DMR 2105191. A portion of this work was performed at the National High Magnetic Field Laboratory (NHMFL), which is supported by National Science Foundation Cooperative Agreements DMR-1644779 and DMR-2128556, and the Department of Energy (DOE). JS acknowledges support from the DOE BES program “Science of 100~T”. The authors declare no competing financial interest. Identification of commercial equipment does not imply recommendation or endorsement by NIST.

\appendix

\section{Appendix A: Sample masses and synthesis}

\begin{table}[h]
\begin{center}
\begin{tabular}{ m{1.3cm}|m{1.7cm}|m{1cm}|m{1.7cm}|m{1.5cm}}
 Sample & Mass (mg) & U:Te & T (°C) & Time \\ 
 \hline
 S1 & 5.9 & 2:3 & 1060/1000 & 1 week\\
 S2 & 1.2 & 5:9 & 900/830 & 2 weeks\\ 
 S3 & 0.7 & 5:9 & 900/830 & 2 weeks\\ 
 S4 & 8.0 & 5:9 & 900/830 & 2 weeks\\ 
 S5 & 4.4 & 5:9 & 900/830 & 2 weeks\\ 
\hline
 P1 & 0.8 & 5:9 & 900/830 & 2 weeks \\ 
 P2 & 1.4 & 2:3 & 900/830 & 2 weeks\\ 
 P3 & 0.4 & 5:9 & 900/830 & 2 weeks\\ 
\end{tabular}
\caption{Sample masses and growth conditions: the initial molar ratio of U to Te (``U:Te"), the temperature gradient used (``T (°C)"), and the length of time the growth was held at those temperatures (``Time").}
\label{tab:samps}
\end{center}
\end{table}

\section{Appendix B: Converting magnetic field angle to magnetic vector components}
\label{sec:app-diags}

\begin{figure}[h!]
    \centering
    \includegraphics[width=0.5\textwidth,trim={1cm 30cm 1cm 2cm},clip]{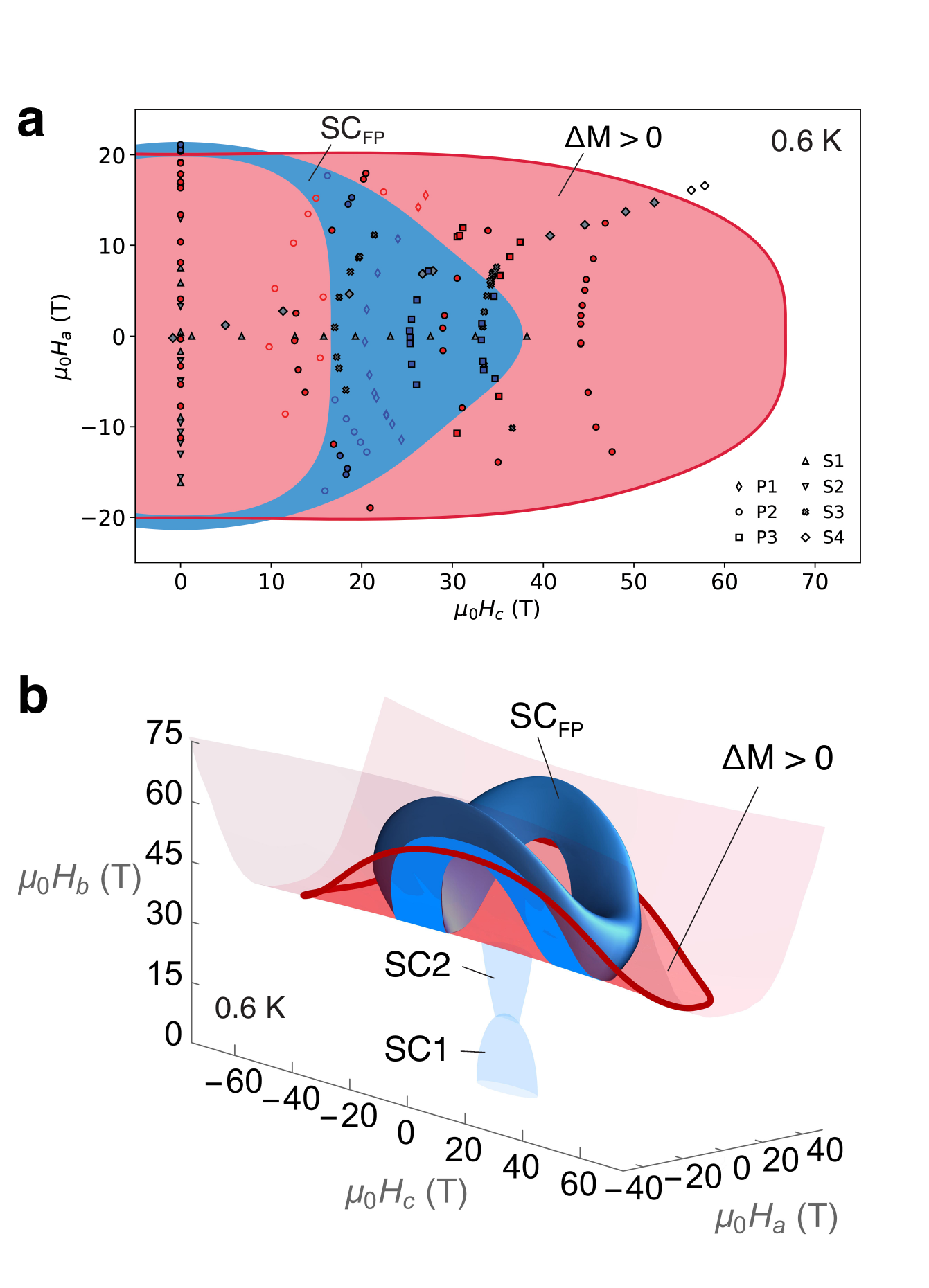}
    \caption{(a) High-field phase diagram for UTe$_2$ at 0.6 K as a function of $H_a$ and $H_c$ based on magnetization and PDO measurements. All symbols are the same as those in Fig. \ref{fig:phasediag}a. (b) An illustration of the 0.6 K phase diagram of UTe$_2$ as a function of magnetic field components, based on the measurements shown in (a) as well as Ref. \cite{LewinFieldangle2024, Lewin2024high}.}
    \label{fig:appphasediag}
\end{figure}

For ease of comparison with other works, Fig. \ref{fig:appphasediag} is a version of Fig. \ref{fig:phasediag} with different axes. Rather than plotting the phase diagram as function of $\theta_a$ and $\theta_{bc}$, we show it here as a function of the magnetic field components along the crystallographic axes: $H_a$, $H_b$, and $H_c$.

\begin{figure}[h!]
    \centering
    \includegraphics[width=0.5\textwidth,trim={6cm 20.5cm 5cm 7.5cm},clip]{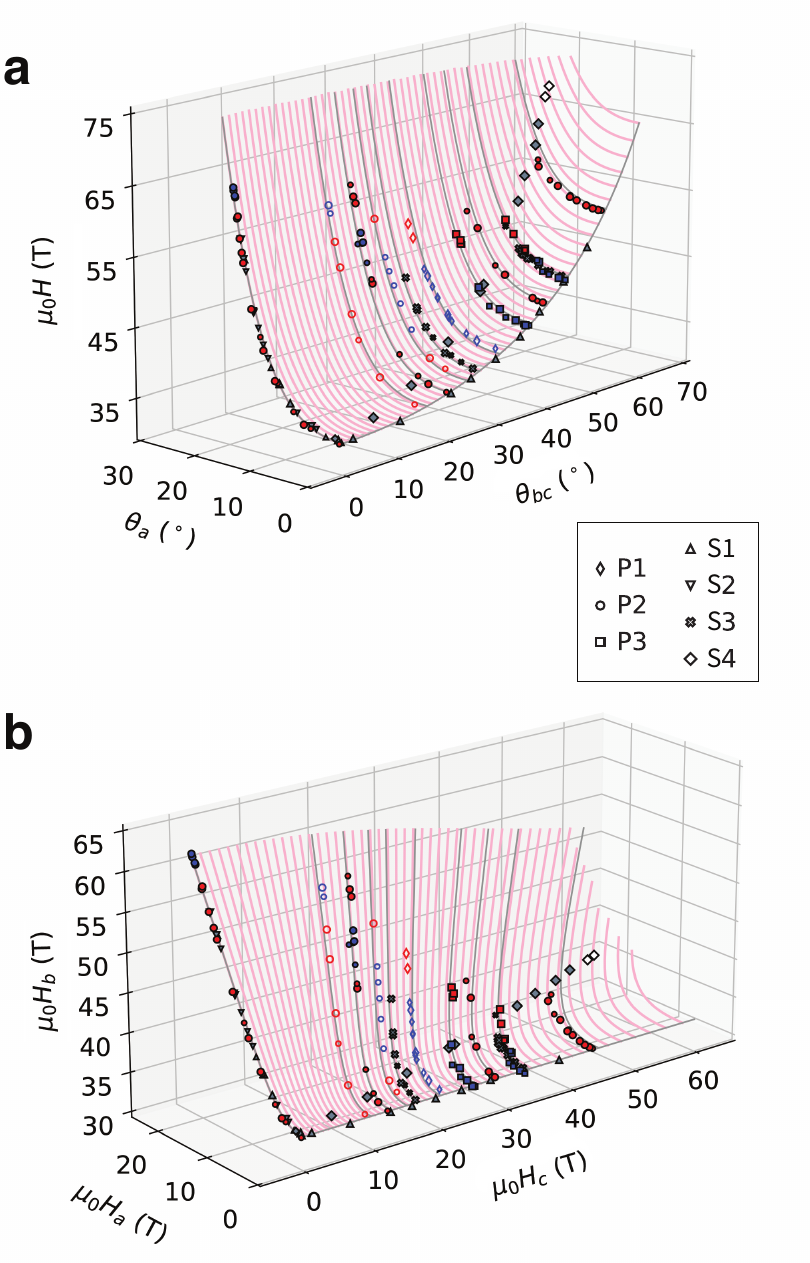}
    \caption{Superconducting and step-like metamagnetic transitions of UTe$_2$ measured by PDO and magnetometry at 0.6 K, shown (a) as a function of $\theta_a$, $\theta_{bc}$, and field strength and (b) as a function of magnetic field components $H_a$, $H_b$, and $H_c$. All symbols are the same as those in Fig. \ref{fig:phasediag}a. For visual clarity, all data are shown at positive $\theta_a$; data points that have been mirrored from negative $\theta_a$ are shown with smaller marker sizes. Pink lines indicate the expected metamagnetic transition at fixed values of $\theta_{bc}$, based on the relationship between $H_m$ and field angle given in Ref. \cite{Lewin2024high}. Gray lines emphasize values of $\theta_{bc}$ at which data were taken.}
    \label{fig:app3dpoints}
\end{figure}

The data points shown in \ref{fig:phasediag}a and \ref{fig:appphasediag}a indicate the points at which either a metamagnetic (red) or superconducting (blue) onset transition was measured by PDO, as well as points at which steps were seen in magnetometry (gray). We show these points in three dimensions in Fig. \ref{fig:app3dpoints}, with axes of field strength and direction in Fig. \ref{fig:app3dpoints}a and with axes of field components in Fig. \ref{fig:app3dpoints}b. These plots correspond to the three-dimensional illustrations in Fig. \ref{fig:phasediag}b and Fig. \ref{fig:appphasediag}b, respectively.

\section{Appendix C: PDO features in the $ab$ plane}
\label{sec:app-PDO}

For certain field sweeps in the $ab$ plane, an increase in PDO frequency can be observed for fields above the metamagnetic transition field; as an example, see the measurement at $\theta_a =$~18.2° in Fig. \ref{fig:slices}a. Additionally, for several field sweeps above 19°, subtle upturns in PDO frequency occur at fields between 55 and 60~T.

An increase in PDO frequency can be caused by a decrease in sample resistance, but this may simply be due to negative magnetoresistance rather than the onset of superconductivity. Changes in PDO frequency can also be due to changes in magnetic moment of the sample being measured \cite{altarawneh2009proximity}.

\begin{figure}[h!]
    \centering
    \includegraphics[width=0.5\textwidth,trim={0cm 0cm 0cm 1cm},clip]{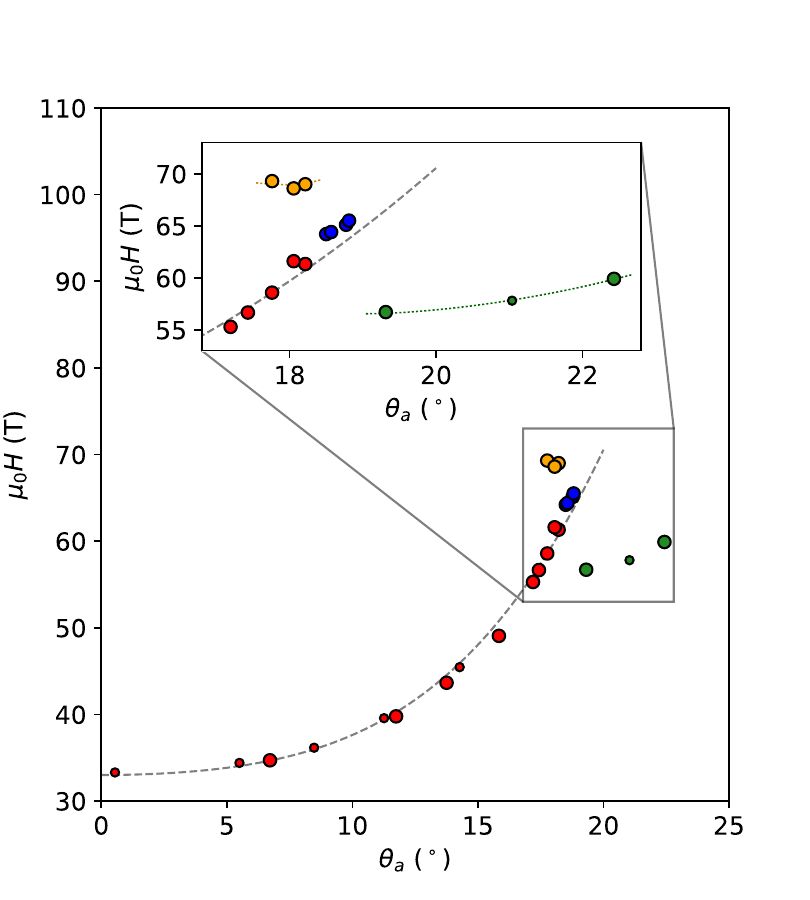}
    \caption{Features observed in the PDO frequency of sample P2 with fields in the $ab$ plane. For visual clarity, all data are shown at positive $\theta_a$; data points that have been mirrored from negative $\theta_a$ are shown with smaller marker sizes. Blue (red) points indicate a superconducting (step-like metamagnetic) transition. Yellow points indicate upturns in PDO frequency at fields above the metamagnetic transition, while green points indicate subtle upturns in PDO frequency observed for $\theta_a >$~19°. The gray dashed line shows the expected evolution of $H_m$ versus $\theta_a$ given in Ref. \cite{Lewin2024high}. Green and yellow dotted lines in the inset are guides to the eye.}
    \label{fig:appPDO}
\end{figure}

We plot the field at which the upturns in PDO frequency occurs as a function of $\theta_a$ in Fig. \ref{fig:appPDO}, along with the metamagnetic and superconducting transitions. It is clear that both the low-angle and the high-angle upturns occur at fields far from the superconducting transitions, even when very close in $\theta_a$. From this, we conclude that these upturns in PDO frequency are unrelated to the SC$_{\rm{FP}}$ phase.

\section{Appendix D: Change in PDO features with increasing $\theta_{bc}$}

\begin{figure}[h!]
    \centering
    \includegraphics[width=0.5\textwidth,trim={0.5cm 0cm 3.5cm 0.5cm},clip]{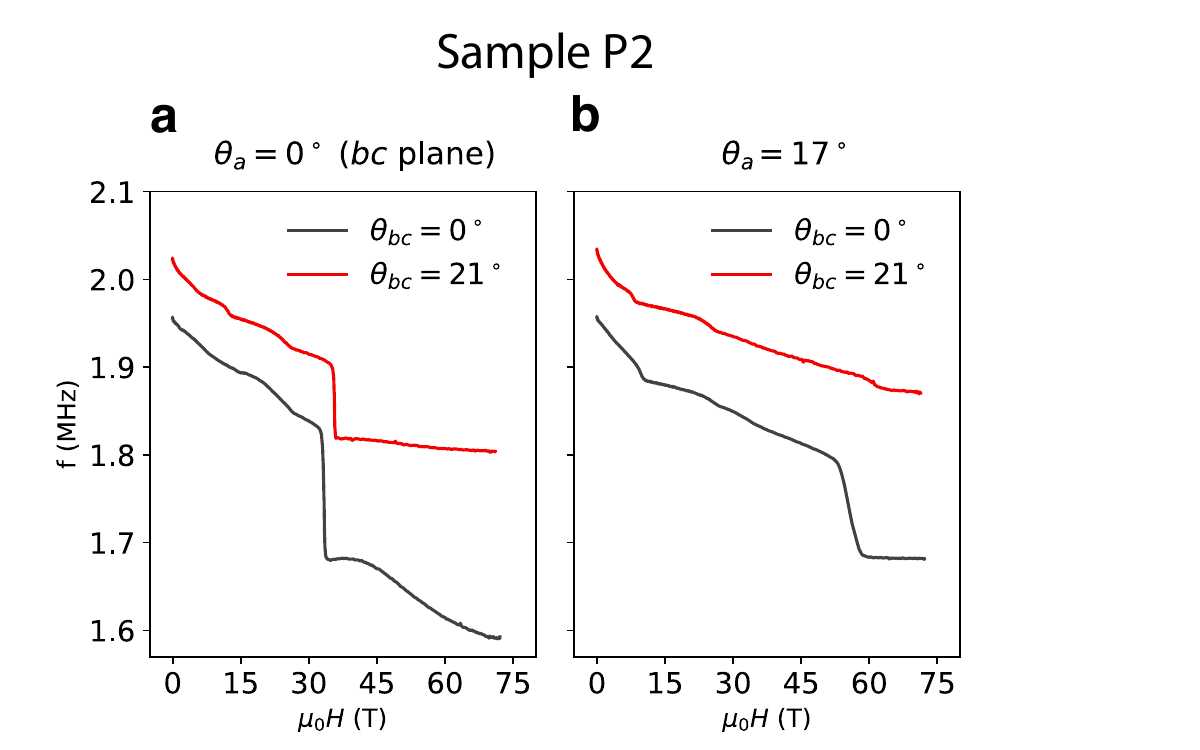}
    \caption{PDO measurements of sample P2 at $\approx$0.65 K at (a) $\theta_{a}$~=~0° and (b) $\theta_a$~=~17°, with each subfigure showing data taken at two different values of $\theta_{bc}$.}
    \label{fig:appPDOth}
\end{figure}

Each dataset is collected at fixed $\theta_{bc}$, using a rotator to change $\theta_a$ between field pulses. However, it is instructive to compare data taken from different datasets at similar $\theta_a$ in order to see trends in the data as a functiom of $\theta_{bc}$. Fig. \ref{fig:appPDOth} shows such PDO data from sample P2. It is clear that without changing $\theta_a$, increasing $\theta_{bc}$ leads to a smaller jump in the PDO signal at the metamagnetic transition. The jump in magnetization at $H_m$ is largest with field along $b$ and decreases both with increasing $\theta_a$ and with increasing $\theta_{bc}$.

\section{Appendix E: Analysis of magnetometry data}
\label{sec:app-M}

\subsection{Determining $\Delta M$}

Our magnetometry measurements include a compensation coil to offset the signal induced by the changing external magnetic field. \textcolor{black}{The compensation coil is designed such that there is no background contribution at room temperature. However, thermal contraction leads to imperfect compensation at cryogenic temperatures.} In many pulsed-field magnetometry measurements, the background remaining after compensation can be determined by taking measurements with an empty coil; this background can then be subtracted from the sample measurement.

Our measurements are taken on a rotator, with the measurement coil rotating along with the sample itself, meaning that the background changes at each rotation angle. In this setup, it is impractical to take empty-coil measurements corresponding to each actual measurement as a method of background subtraction. Therefore, we do have non-zero backgrounds to contend with, sometimes varying as a function of field.

\begin{figure}[h!]
    \centering
    \includegraphics[width=0.5\textwidth,trim={0cm 0cm 1cm 1cm},clip]{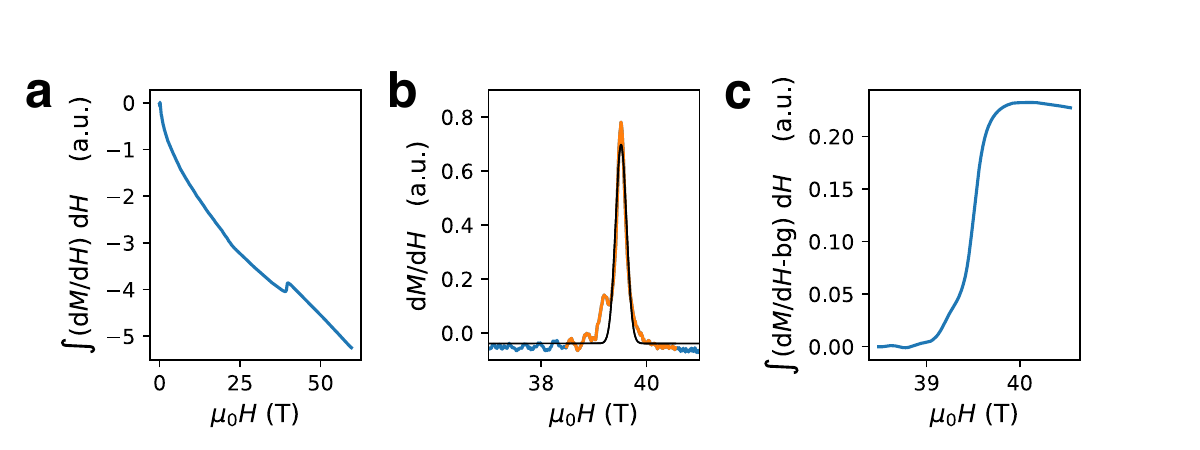}
    \caption{An example of the analysis process for magnetometry data from a single field sweep. (a) The result of integrating $\mathrm{d}M/\mathrm{d}H$ across the full field range with no background subtraction. (b) The blue curve is $\mathrm{d}M/\mathrm{d}H$ as measured. The black line shows the best-fit Gaussian with constant background, as described in the text. The orange subsection of $\mathrm{d}M/\mathrm{d}H$ shows the region between $H_{-}$ and $H_{+}$ as described in the text. (c) The result of integrating $\mathrm{d}M/\mathrm{d}H$ minus its constant background, from $H_{-}$ to $H_{+}$.}
    \label{fig:appdeltas}
\end{figure}

As described in the main text, we extract $\mathrm{d}M_i/\mathrm{d}H$ up to a constant of proportionality from our measurements. If we had no background, we could integrate this to find something proportional to $M_i$ as a function of field. Because of the background, we measure additional contributions that are not related to the sample. An example of the result of integrating what we call $\mathrm{d}M_i/\mathrm{d}H$ is shown in Fig. \ref{fig:appdeltas}a. The metamagnetic transition can be seen as a sudden rise around 40~T, but the negative slope is non-physical.

In the region around the metamagnetic transition, the background of $\mathrm{d}M_i/\mathrm{d}H$ can be approximated as constant (it can be seen that the result of integration is roughly linear). Our goal is to subtract this background and then to integrate $\mathrm{d}M_i/\mathrm{d}H$ over a small region around the metamagnetic transition. We fit $\mathrm{d}M_i/\mathrm{d}H$ to a Gaussian with a constant background, of the form $a \exp{-\frac{(H-H_m)^2}{2\sigma ^2}} + \textrm{bg}$. Here $a$, bg, and $\sigma$ are constants and $H_m$ is the metamagnetic transition field (also a parameter in the fit). An example of such a fit is shown in Fig. \ref{fig:appdeltas}b. In order to calculate $\Delta M$, we integrate $\mathrm{d}M_i/\mathrm{d}H$ - bg from $H_{-}$ to $H_{+}$. For the majority of our analysis, we use $H_{\pm} = H_m \pm 6 \sigma$; for exceptionally sharp transitions we use $H_{\pm} = H_m \pm 10 \sigma$. An example of this integration is shown in Fig. \ref{fig:appdeltas}c.
In cases where $dM/dt$ itself had a clear polynomial background, usually due to inadequate compensation, we subtracted the polynomial background before doing the analysis described above.

\subsection{Hysteresis}

\begin{figure}[h!]
    \centering
    \includegraphics[width=0.5\textwidth,trim={-1cm 0cm -2cm 0.5cm},clip]{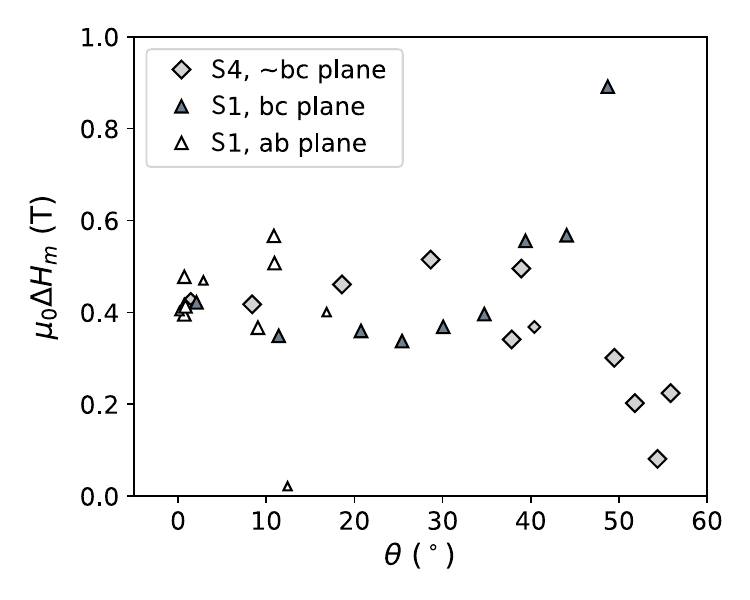}
    \caption{The width of the hysteresis loop for samples S1 and S4 as a function of field angle away from the $b$ axis, calculated as described in the text.}
    \label{fig:apphyst}
\end{figure}

We define $\Delta H_m$ as the difference in the field at which the metamagnetic transition occurs in field upsweeps vs. field downsweeps, i.e. the width of the hysteresis loop. For each field sweep, we fit $\mathrm{d}M_i/\mathrm{d}H$ in the vicinity of the metamagnetic transition to a Gaussian centered about $H_m$, as described above. We use the values of $H_m$ extracted from this fit for upsweeps and downsweeps at the same field angle to find $\Delta H_m$. The results for samples S1 and S4 are shown in Fig. \ref{fig:apphyst}.

For fields near the $b$ axis, both samples have $\mu_0\Delta H_m \approx$~0.4~T, consistent with previous results at low temperature \cite{miyake2019metamagnetic,LewinFieldangle2024}. 

For sample S1 tilting in the $bc$ plane and sample S4 tilting approximately in the $bc$ plane, the width of the hysteresis loop is steady up to $\theta_a \approx$~40°. For higher angles, the size of $\Delta H_m$ for sample S1 increases with angle while the size of $\Delta H_m$ for sample S4 decreases with angle.

For sample S1 in the $ab$ plane, the overall trend as a function of field angle is that the width of the hysteresis loop is steady up to $\theta_a^{crit}$, the highest angle at which a metamagnetic transition could be observed in magnetometry. However, there is one field angle at which $\Delta H_m$ is near zero.

Both this outlier in the $ab$ plane and the contradictory behavior of S1 and S4 in the $bc$ plane indicate that these measurements of $\Delta H_m$ may not be entirely accurate. Sample heating is a possible cause of these discrepancies. In pulsed-field measurements, samples experience some amount of heating due to eddy currents. In the magnets in which these measurements were taken, this heating is primarily on the upsweep, which has a higher rate-of-change of magnetic field than the downsweep. Given that $H_m$ shifts slightly with temperature, measurements of $\Delta H_m$ can be affected by changes in sample temperature. The amount of heating a sample experiences depends not only on its mass, but also on the area of the sample face perpendicular to the applied field. Thus, it is possible for the same sample to experience slightly different amounts of heating at different field angles. 

Our measurements of $\Delta H_m$ indicate that the metamagnetic transition is first-order up to $\theta_a^{crit}$ in the $ab$ plane and up to at least 50° in the $bc$ plane. However, due to the discrepancies described above, we have not relied on these measurements to draw conclusions about the order of the metamagnetic transition.

\subsection{Data from Sample S3: rotations in $\theta_{a}$ at finite $\theta_{bc}$}

\begin{figure}[h!]
    \centering
    \includegraphics[width=0.5\textwidth,trim={0cm 0cm 0cm 0cm},clip]{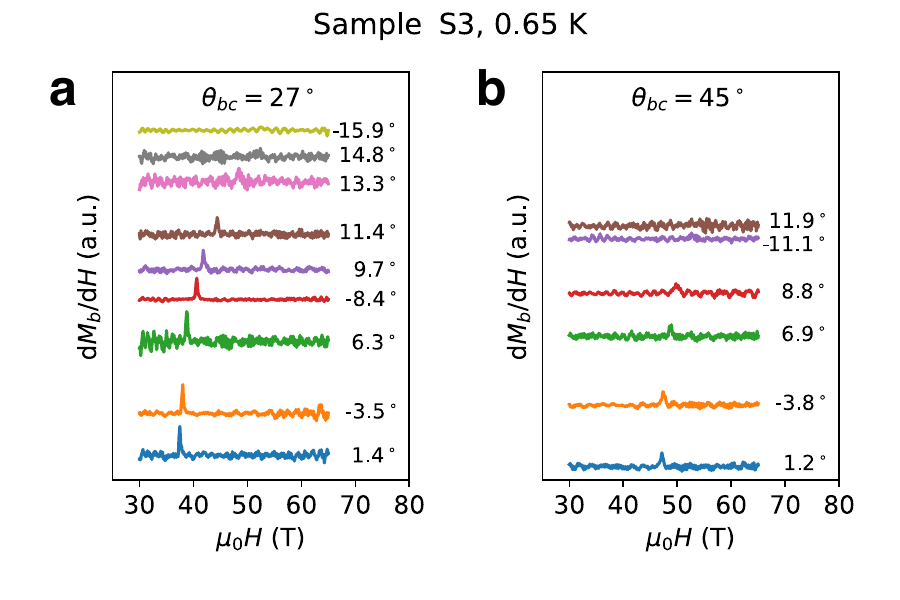}
    \caption{Measurements of $\mathrm{d}M_b/\mathrm{d}H$ versus field strength for sample S3 with field pulses taken at various $\theta_a$ for (a) $\theta_{bc}$~=~28° and (b) $\theta_{bc}$~=~45°.}
    \label{fig:appS3}
\end{figure}

Fig. \ref{fig:appS3} shows differential susceptibility measurements of sample S3. Those in Fig. \ref{fig:appS3}b are a representative subset of the measurements that were used in constructing Fig. \ref{fig:slices}d.

The signal-to-noise ratio for the measurements in Fig. \ref{fig:appS3} is clearly lower than for those in Fig. \ref{fig:mag}. The weaker signal can be attributed to S3's lower mass compared to the other magnetometry samples, as well as the large angle between the applied field and the $b$ axis for all of the measurements on S3. The oscillatory features in the data are caused by time-periodic mechanical oscillations in the measurement coil.

\subsection{Data from Sample S5: $c$-axis magnetization}

\begin{figure}[h!]
    \centering
    \includegraphics[width=0.5\textwidth,trim={2cm 1cm 1cm 0cm},clip]{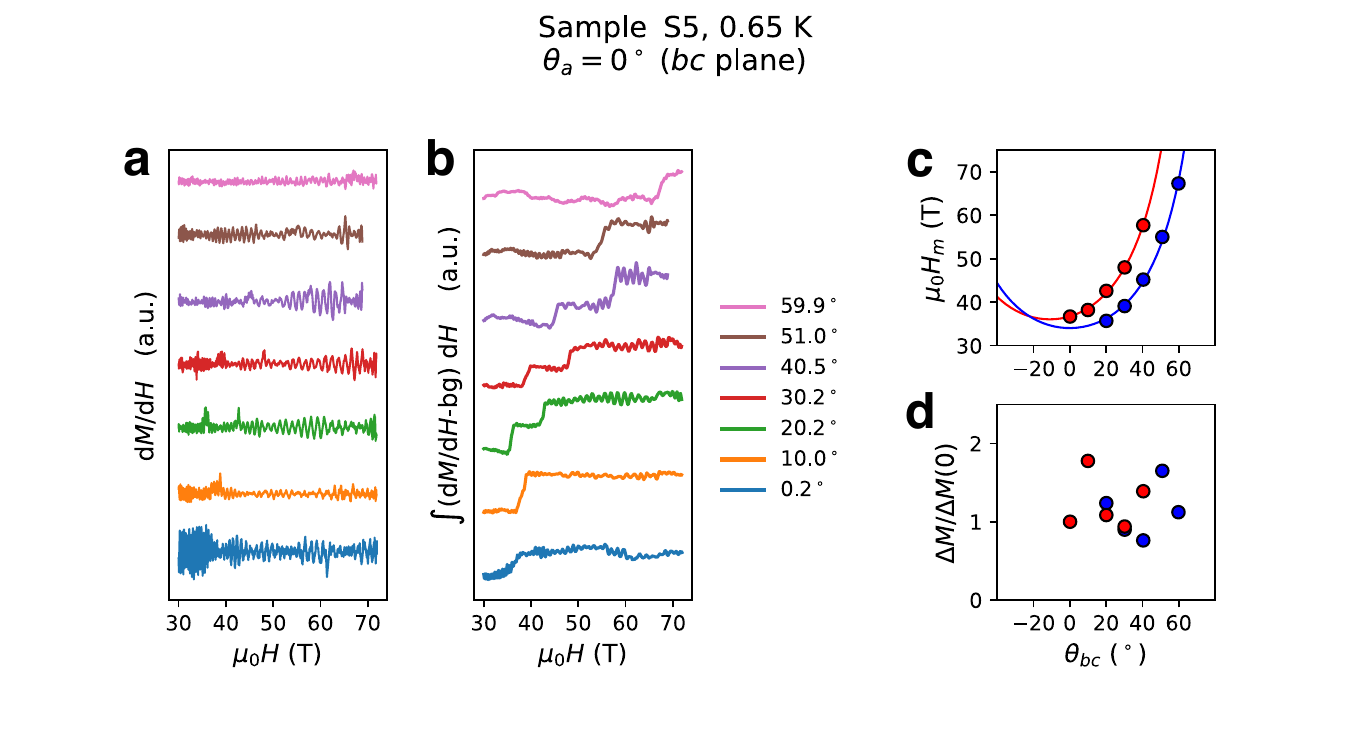}
    \caption{(a) Measurements of $\mathrm{d}M/\mathrm{d}H$ versus field strength for sample S5 with field pulses at various angles $\theta_{bc}$ within the $bc$ plane. (b) The result of integrating the background-subtracted $\mathrm{d}M/\mathrm{d}H$ curves. (c) Fields at which steps in magnetization occur in sample S5, as a function of $\theta_{bc}$. Solid lines are fits to $H_m^b/ \cos({\theta_{bc}})$, the expected evolution of $H_m$ in the $bc$ plane. (d) The size of the steps in magnetization of sample S5 as a function of $\theta_{bc}$,  normalized to the size of the step at $\theta_{bc}\approx$~0. }
    \label{fig:appM6}
\end{figure}

Sample S5 was measured with the magnetometer coil aligned with the sample $c$ axis and with field in the $bc$ plane ($\theta_a$~ =~0). Our measurements indicate that $M_c$ is not affected by the metamagnetic transition.

Fig. \ref{fig:appM6}a shows differential susceptibility measurements of sample S5 at various values of $\theta_{bc}$, where $\theta_{bc}$~=~0 corresponds to field along the $b$ axis. The signal from this sample is barely distinguishable above the noise; compare to the data shown in Fig. \ref{fig:mag}. We conclude that there is no change in $M_c$ at the metamagnetic transition, and that the observed signal is due to minor sample misalignment with the coil, i.e., a small projection of the sample $b$ axis along the coil.

Fig. \ref{fig:appM6}b shows the integration of the differential susceptibility curves with constant background subtracted, using the analysis technique described above. Fig. \ref{fig:appM6}c shows the field value at which the steps in magnetization occur in this sample, as a function of $\theta_{bc}$. It can be seen that there are two steps in the sample magnetization for certain values of $\theta_{bc}$. It appears that rather than a single crystalline domain, sample S5 is comprised of two domains that are not aligned with each other. The solid lines in Fig. \ref{fig:appM6}c show fits to the expected evolution of $H_m$ in the $bc$ plane for both transitions. The results indicate that both domains are rotating more or less in the $bc$ plane.

Fig. \ref{fig:appM6}d shows the size of the change in magnetization of the sample as a function of $\theta_{bc}$. The data are normalized to the size of the step at $\theta_{bc}\approx$~0. There is not a clear trend in step size as a function of angle, though the scatter is quite large, attributed to the small signal-to-noise ratio of the data. If the observed signal were due to $M_c$, then $\Delta M$ should be 0 with field along the $b$ axis due to the orthorhombic symmetry of UTe$_2$.

\subsection{Integrated susceptibility}

\begin{figure*}[h!]
    \centering
    \includegraphics[width=\textwidth,trim={2cm 1cm 0.5cm 1cm},clip]{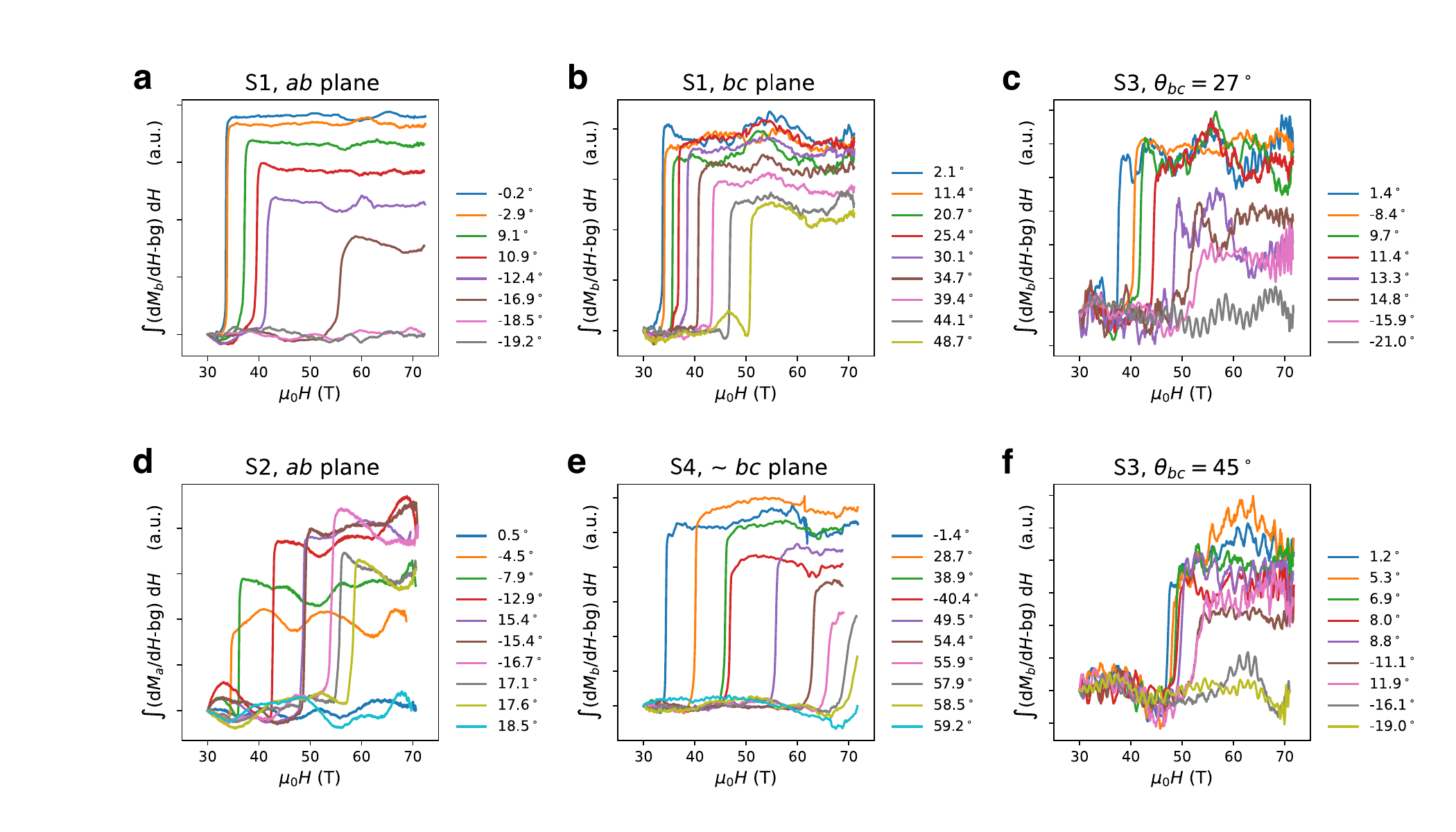}
    \caption{Integration of representative differential susceptibility measurements for samples S1, S2, S3, and S4 with constant background subtracted. (a) Sample S1 at various $\theta_a$ in the $ab$ plane. (b) Sample S1 at various $\theta_{bc}$ in the $bc$ plane. (c) Sample S3 at various $\theta_{a}$ for fixed $\theta_{bc}$~=~28°. (d) Sample S2 at various $\theta_a$ in the $ab$ plane. (e) Sample S4 at various $\theta$ from the $b$ axis, tilting in a plane approximately 10° rotated from the $bc$ plane. (f) Sample S3 at various $\theta_{a}$ for fixed $\theta_{bc}$~=~45°. }
    \label{fig:appInt}
\end{figure*}

Fig. \ref{fig:appInt} shows the integration of differential susceptibility for samples S1, S2, S3, and S4 with constant background subtracted, using the analysis technique described above. Data from representative angles are shown for each sample. From the measurements of sample S4 shown in Fig. \ref{fig:appInt}e, it can be seen that a step in magnetization occurs up to at least 58.5° away from the $b$ axis. For 57.9° and 58.5°, it appears that the maximum measurement field of 73~T was not enough to reach the upper field of the metamagnetic transition, so that only a partial step was measured. For 59.2° it appears possible that a metamagnetic transition onsets just below 73~T, but we cannot draw any definitive conclusions.

\begin{figure*}[h!]
    \centering
    \includegraphics[width=\textwidth,trim={2cm 1cm 0.5cm 1cm},clip]{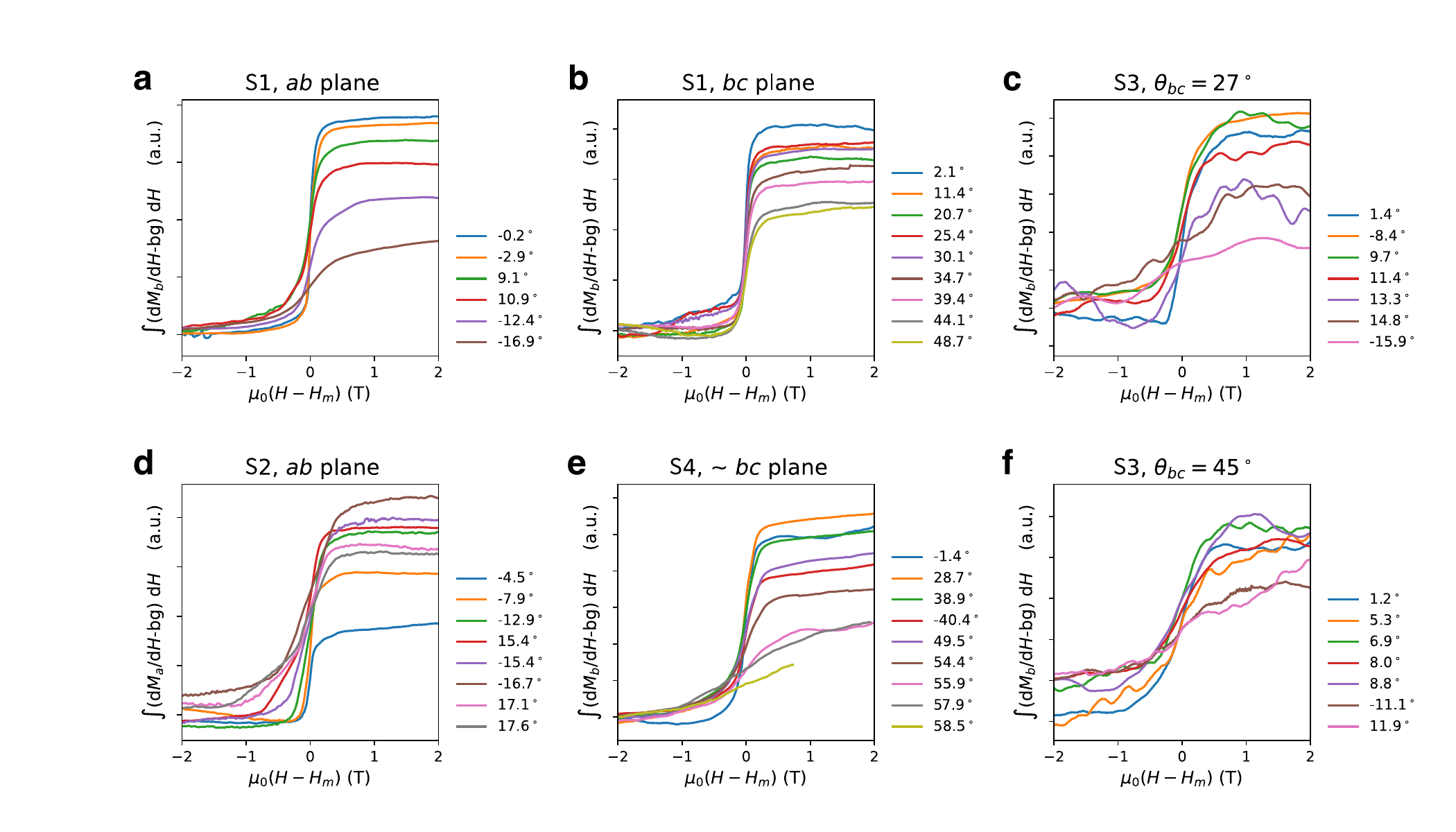}
    \caption{All of the curves from Fig. \ref{fig:appInt} that show a metamagnetic transition are replotted here with the $x$ axis shifted in order to compare transition widths.}
    \label{fig:appIntShift}
\end{figure*}

In order to compare the transition widths of the metamagnetic transition, it is helpful to shift the integrated susceptibility curves shown in Fig. \ref{fig:appInt}. For each field sweep in that figure for which a metamagnetic transition was observed, we have replotted the integrated susceptibility shifted by $H_m$, the field of the metamagnetic transition. The results are shown in Fig. \ref{fig:appIntShift}. We observe that there is a gradual increase in the width of the transition as field is tilted away from the $b$ axis (Fig. \ref{fig:appIntShift}a,b,d,e) and as field is tilted out of the $bc$ plane (Fig. \ref{fig:appIntShift}c,f). This makes it difficult to use the transition width to uniquely distinguish first-order from continuous transitions in this dataset.

\section{Appendix F: Uncertainty estimates}

Each sample is hand-mounted on the single axis rotator described in the Methods section. In the case of rotations in the $ab$ ($bc$) plane, the sample is aligned with its $c$ ($a$) axis along the axis of rotation. In the case of rotations at fixed $\theta_{bc}$, the sample is mounted as described in Ref. \cite{Lewin2024high}, with the $a$ axis normal to the rotator platform and the sample's $c$ axis at an angle $\theta_{bc}$ from the axis of rotation.

The alignment of the sample mounting on the rotator is determined by eye. In order to determine the sample's mounting orientation beyond visual inspection, we take advantage of the field-angle evolution of $H_m$ determined in Ref. \cite{Lewin2024high}:

\begin{equation}
    H_m(\theta_{bc},\theta_a) = \frac{H_m^b}{\cos(\theta_{bc})} + \alpha_2 \sin^2 \theta_a + \alpha_4 \sin^4 \theta_a,
\label{eq:Hm}
\end{equation}
where $H_m^b$ is the value of $H_{\rm{m}}$ for \textbf{H}~$\parallel$~$b$. In Ref. \cite{Lewin2024high}, the constants $\alpha_2$ and $\alpha_4$ were determined by fitting to be $\alpha_2 = 95$ T, $\alpha_4 = 1934$ T.

The angle-dependence of $H_m$ for sample S1 when rotated in the $ab$ and $bc$ planes was consistent with Eq. \ref{eq:Hm}, confirming the sample was mounted in the intended orientations. The same was true for sample P2 rotated in the $ab$ plane. For sample S4, rotations that were intended to be in the $bc$ plane yielded values of $H_m$ that rose more rapidly with angle than the $1/\cos(\theta)$ dependence expected for rotations in the $bc$ plane. By fitting the angle-dependence of $H_m$ to Eq. \ref{eq:Hm}, we determined that sample S4 was likely rotating $\approx$ 13° off the $bc$ plane.

For samples mounted at fixed $\theta_{bc}$ and being rotated out of the $bc$ plane, $H_m$ is at a minimum when the field is in the $bc$ plane and increases with increasing $\theta_a$. The minimum value of $H_m$ for a given dataset will be $H_m^0~\equiv~\nicefrac{H_m^b}{\cos(\theta_{bc})}$. Therefore
\begin{equation}\label{eq:thbc}
\theta_{bc} = \arccos\left(\frac{H_m^b}{H_m^0}\right).
\end{equation}
For most datasets, $H_m^0$ was measured and can be seen by symmetry to be the minimum value of $H_m$ versus $\theta_a$. For datasets in which measurements were not taken in or near the $bc$ plane, the value of $H_m^0$ was determined by fitting the evolution of $H_m$ vs $\theta_a$ to Eq. \ref{eq:Hm}.

Extracting $\theta_{bc}$ from Eq. \ref{eq:thbc} requires using the value of $H_m^b$, which carries some uncertainty if $H_m^b$ has not been measured for the sample. Propagating error from Eq. \ref{eq:thbc} yields
\begin{equation}\label{eq:appth}
\Delta \theta_{bc} = \frac{-1}{\sqrt{(H_m^0)^2-(H_m^b)^2}} \cdot \Delta H_m^b.
\end{equation}
It is clear that our error in estimating $\theta_{bc}$ is greatest for $H_m^0$ near $H_m^b$; that is, for fields near the $b$ axis.

In Ref. \cite{Lewin2024high} the measurements on sample P2 were calculated to be at $\theta_{bc}$~=~8° and 23° based on an estimated $H_m^b = 34$~T. In this work, we measured sample P2 in the $ab$ plane and found $H_m^b \approx 33$~T. With this measured value of $H_m^b$ for sample P2, the $\theta_{bc}$ at which previous measurements were taken are recalculated to be $\theta_{bc}$~=~16° and 26°. These updated values are reflected in all of the plots in this paper.

Samples P1, P3, and S3 were measured at fixed $\theta_{bc}$ and without measured $H_m^b$. For these samples we estimate $H_m^b = 34$~T to calculate $\theta_{bc}$. For sample P1, $H_m^0$ was measured to be $\approx$~39~T. Based on Eq. \ref{eq:appth}, even if our estimate of $H_m^b$ is off by 1~T the error in calculating $\theta_{bc}$ is at most $\pm$~3° for this dataset of sample P1. Samples P3 and S3 were measured at angles with even higher $H_m^0$, so $\Delta \theta_{bc} \leq$~3° for all of these measurements.

All of the above uncertainty is related to calculation of a sample's hand-mounted position on the rotator. The uncertainty is much less when considering the angles of the rotator itself. As described in Ref. \cite{Willis2020goniometer}, the rotator is equipped with a tilt coil used to measure its angular position and the precision of the rotator is $\pm$~0.2°. The sample is often offset relative to the tilt coil in terms of the rotation axis, but high-symmetry planes are used to fix this offset. For example, a sample at fixed $\theta_{bc}$ rotated in $\theta_a$ plane should have symmetric data about the $bc$ plane. Its relative rotation for each pulse is determined via the tilt coil and its absolute rotation is found by requiring that the data be symmetric about $\theta_a = 0$.


\bibliography{magbib}


\end{document}